\documentclass[journal=jacsat,manuscript=article]{achemso}
 \usepackage[version=3]{mhchem} % Formula subscripts using \ce{}
 \usepackage{amsmath,bm}
 \usepackage{mathrsfs}
 \usepackage{graphicx}
 \usepackage{setspace} %leaves captions single space in draft mode
 \usepackage{epstopdf}
 \usepackage{dcolumn}
 \usepackage{amsmath}
 \usepackage{epsfig}
 \usepackage{indentfirst}
 \usepackage{psfrag}
 \usepackage{subfigure}
 \usepackage{color}
 \usepackage{units} % rz
\usepackage{enumitem}

\usepackage{physics}
\usepackage{dcolumn}% Align table columns on decimal point
\usepackage{bm}% bold mathhttps://www.overleaf.com/project/5af95d0f3775594d14d4a052

\newcommand{\Figref}[1]{Fig.~\ref{#1}}

\usepackage{ulem}[normalem] %Whatch out, places underlining in Journal references. Use \normalem before references to deactivate this 

\newcommand{\np}[1]{\textcolor{black}{ #1}}

\makeatletter
\newcommand\colorsout[1]{\bgroup \markoverwith{\textcolor{#1}{\rule[0.5ex]{2pt}{0.4pt}}}\ULon}

\makeatother

\usepackage{xr-hyper}% To include reference from the supplementary information, which is an external file: "Supplementary.tex"

\usepackage[backref=none,bookmarksnumbered=true,bookmarks=true,bookmarksopen=true,colorlinks=true,
citecolor=blue,linkcolor=blue,anchorcolor=green,urlcolor=blue,unicode=false]{hyperref}
\makeatletter
\newcommand*{\addFileDependency}[1]{% argument=file name and extension
  \typeout{(#1)}
  \@addtofilelist{#1}% For OverLeaf to compile external supp info files
  \IfFileExists{#1}{}{\typeout{No file #1.}}
}
\makeatother

\newcommand*{\myexternaldocument}[1]{%
    \externaldocument{#1}%
    \addFileDependency{#1.tex}% For OverLeaf to compile external supp info files
    \addFileDependency{#1.aux}%
}

\myexternaldocument{Supplementary}% Include the supplementary reference (Figure label). Here should not specify the extension .tex

\graphicspath{ {Fig/} {Figsi/} }  %  Look in the folder Fig and Figsi to upload figures 

\usepackage{miller}

\newcommand{\nanogune}{CIC nanoGUNE-BRTA, 20018 Donostia-San Sebasti\'an, Spain}
\newcommand{\dipc}{Donostia International Physics Center (DIPC), 20018 Donostia-San Sebastián, Spain}
\newcommand{\cfm}{Centro de Física de Materiales MPC (CSIC/UPV-EHU), 20018 Donostia-San Sebastián, Spain}
\newcommand{\ikerbasque}{Ikerbasque, Basque Foundation for Science, 48013 Bilbao, Spain}

\author{Jeremy Hieulle}
\affiliation{\nanogune}
\email{jeremy.hieulle@uni.lu}

\author{Carlos Garcia Fernandez}
\affiliation{\dipc}

\author{Niklas Friedrich}
\affiliation{\nanogune}

\author{Alessio Vegliante}
\affiliation{\nanogune}

\author{Sofia Sanz}
\affiliation{\dipc}

\author{Daniel Sánchez-Portal}
\affiliation{\dipc}
\alsoaffiliation{\cfm}

\author{Michael M. Haley}
\affiliation{Department of Chemistry \& Biochemistry and the Materials Science Institute, University of Oregon, Eugene, Oregon 97403-1253, United States}

\author{Juan Casado}
\affiliation{Department of Physical Chemistry, University of Malaga, Campus de Teatinos s/n, 229071 Malaga, Spain}

\author{Thomas Frederiksen}
\affiliation{\dipc}
\alsoaffiliation{\ikerbasque}

\author{Jos\'e Ignacio Pascual}
\affiliation{\nanogune}
\alsoaffiliation{\ikerbasque}
\email{ji.pascual@nanogune.eu}

\title{From Solution-to-Surface: Persistence of the Diradical Character of a Diindenoanthracene Derivative on a Metallic Substrate}

\begin{document}
\newpage

\begin{abstract}
Organic diradicals are envisioned as elementary building blocks for designing a new generation of spintronic devices and have been used in constructing prototypical field effect transistors and non-linear optical devices. Open-shell systems, however, are also reactive, thus requiring design strategies to “protect” their radical character from the environment, especially when they are embedded into solid-state devices. Here, we report the persistence on a metallic surface of the diradical character of a diindeno[\textit{b,i}]anthracene (DIAn) core protected by bulky end-groups. Our scanning tunneling spectroscopy measurements on single-molecules detected singlet-triplet excitations that were absent for DIAn species packed in assembled structures. Density functional theory simulations unravel that molecular geometry on the metal substrate can crucially modify the value of the single-triplet gap via the delocalization of the radical sites. The persistence of the diradical character over metallic substrates is a promising finding for integrating radical-based materials in functional devices. 

\end{abstract}

%\begin{tocentry}[h]
\begin{center}

\includegraphics[width=5cm]{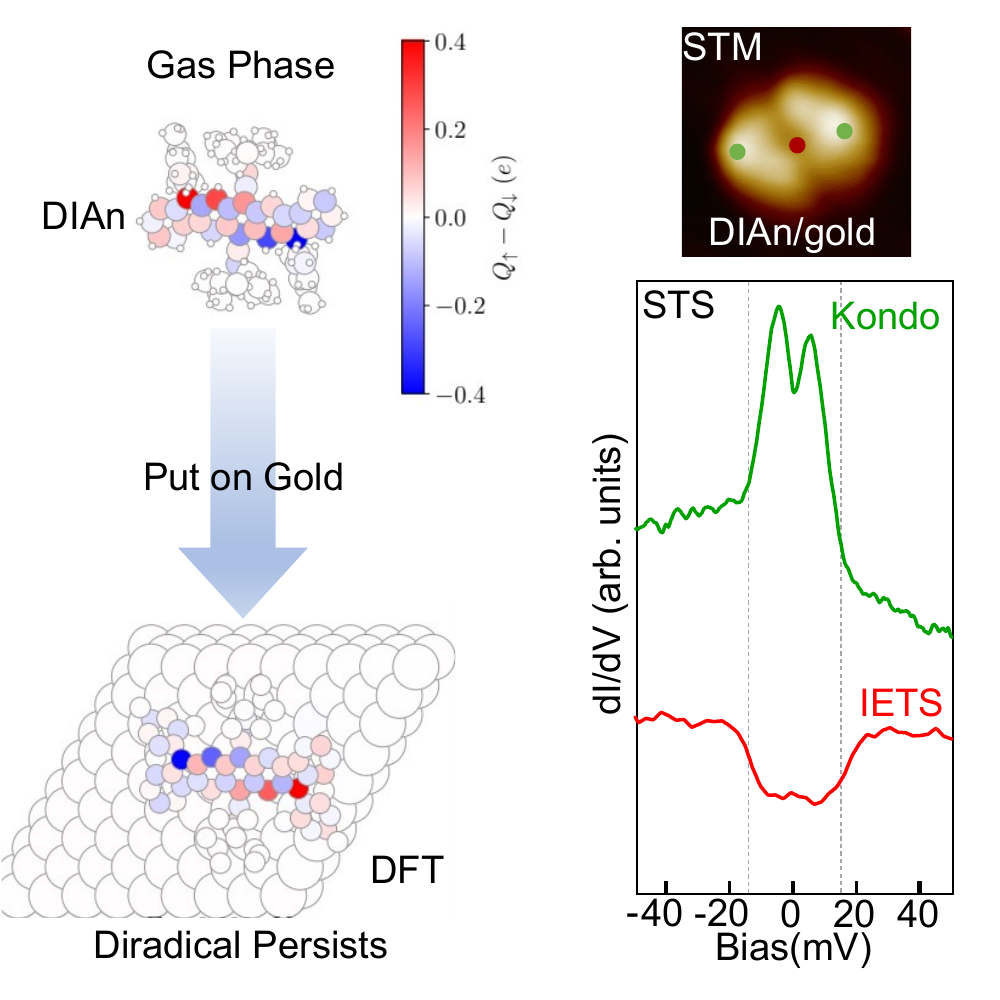}
%  TOC Graphic 
\end{center}
%\end{tocentry}
\maketitle

\newpage
%\section{Introduction}
Diradical molecules are promising candidates for the design of new generation nonlinear optic (NLO) \cite{Fukuda2015} and spintronic devices \cite{Shil2015, Tagami2004}. As examples, diradicals have been used to build spin filters \cite{Shil_2015_Spin_filters}, field-effect transistors\cite{Zong_diradicat_FET}, and they could potentially be used for memory storage devices \cite{Fukuda2015, Tagami2004}. In recent years, much focus has been directed to the understanding of the interactions established between the two spin centers of a diradical, and how those interactions can be controlled \cite{Fukuda2015, Rajca2007, Majzik2018, Karafiloglou1989}.

To date, the impact of a contact metal electrode on the electronic and magnetic properties of diradicals remains unclear, especially since the influence of molecular geometry, adsorption, and packing has yet to be revealed. Those parameters have been shown to play a crucial role in defining the magnetic properties of metal-organic molecules on surfaces \cite{Saiful2020, Zhang2015, Amokrane2017, Karan2016, SanchezGrande2020, DiLullo2012}, and are also expected to affect the properties of purely organic diradicals. Most of the recent studies on diradicals are performed in soft solution or solid environments. Therefore, exploring their structures on metallic surfaces can be of great significance to disentangle remaining unknown questions about these fascinating molecules.

Among all diradical molecules, the diindeno[\textit{b,i}]anthracene (DIAn) framework is of particular interest due to its high solubility in common organic solvents, high thermal stability, ease of sublimation, and excellent oxidation resistance \cite{Rudebusch2016}. DIAn has also exceptional electronic properties, such as a small singlet-triplet gap and a balanced ambipolar charge-transport behavior \cite{Rudebusch_JACS_2016, Dressler2018, DRESSLER_2020, Hayashi_JACS_2020}. Owing to these properties, DIAn molecules are perfectly suitable for mass production for future market applications in, for instance, high-performance organic field-effect transistors \cite{Zong_diradicat_FET}. In a more fundamental scenario, DIAn is ideally suited to investigate, at the atomic scale, the magnetic behavior of prototypical diradical molecules in contact with inorganic interfaces, seeking strategies to maintain their molecular functionality when forming part of solid state devices. 

Here we demonstrate that individual DIAn molecules maintain their diradical character on a metallic substrate, with a reduced singlet-triplet gap compared to values observed in the solid state. We studied DIAn molecules and assemblies adsorbed on an atomically clean Au(111) substrate, and probed inter- and intra-molecular spin interactions by scanning tunneling spectroscopy (STS). \np{Using the Kondo effect as a fingerprint of molecular magnetism, \cite{Ternes2009, Fernandez_Torrente08, Li2019, Li2020, Wang2022, Turco2023, Vilas2023} we probe local spin interactions inside the diradicals, unraveling underlying mechanisms at the origin of magnetic interactions between the two spin centers of the diradical. More importantly, we demonstrate that those magnetic interactions depend on molecular conformation over the surface and on their interaction with other species in assembled nanostructures. In particular, supported by density functional theory (DFT) simulations we show that the rotation of mesityl substituents modifies the magnetic interaction between the two spin centers of a single DIAn molecule. These results outline the relevance of intermolecular interactions in supramolecular nanoarchitectures in the magnetic properties of open-shell molecular systems.}

%\section{Results and discussion}

\begin{figure*}[!ht]
    \includegraphics[width=0.90\textwidth]{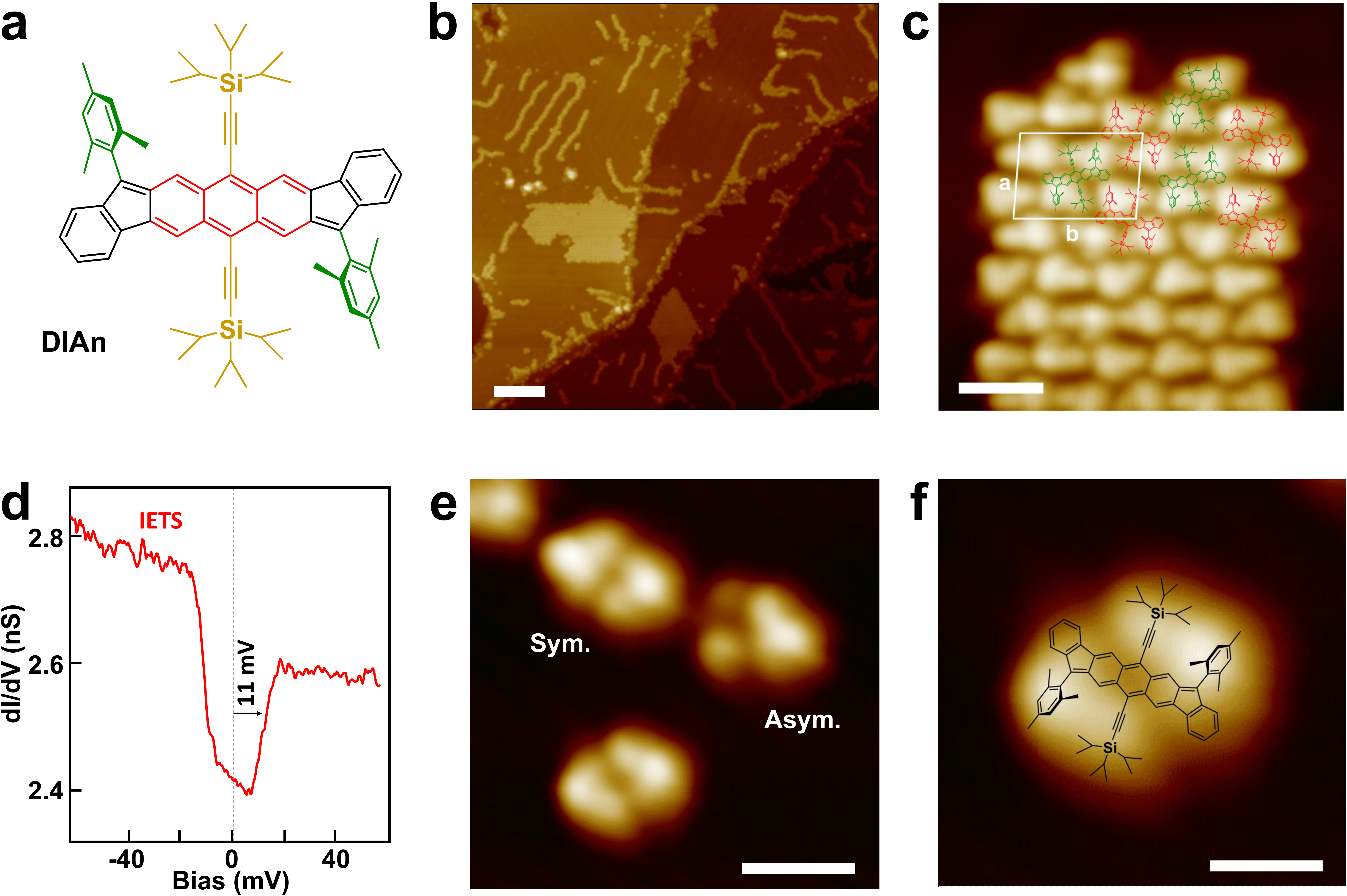}
	\caption{\label{FigTopoDomain} 
	\textbf{Molecular packing of DIAn diradicals on Au(111).} (a) Chemical structure of DIAn. The anthracene core is highlighted in red, while the (triisopropylsilyl)ethynyl groups and mesityl groups are highlighted in orange and green, respectively. Indene groups are drawn in black color. (b) STM image showing the typical molecular chains and close-packed domains formed by DIAn. (c) STM image of the close-packed domain. A model of the molecular packing indicating the bimolecular unit cell is superimposed on the topographic image. (d) $dI/dV$ spectra obtained at the anthracene core of a single diradical molecule in the domain, showing a vibrational inelastic electron tunneling (IETS) feature. (e-f) Topographic images of isolated DIAn monomer obtained by lateral manipulation with the STM tip.  Scale bars: b) 20 nm; c-e) 2 nm; f) 1 nm.}
\end{figure*}

The core of the DIAn molecule is composed of two indene groups (black section, \Figref{FigTopoDomain}a), spaced by an anthracene bridge (red section in \Figref{FigTopoDomain}a). The indene moieties endow DIAn with the radical character of the molecule \cite{Rudebusch2016} and, here, incorporate bulky mesityl substituents (green section in \Figref{FigTopoDomain}a) for protecting them. 
Additionally, (triisopropylsilyl)ethynyl groups (CCSi(i-Pr)$_3$, orange section, \Figref{FigTopoDomain}a) are attached to the quinoidal anthracene core to provide high stability against oxidation and high solubility. Details of the synthesis of DIAn have been reported previously \cite{Rudebusch2016}.

We deposited DIAn diradicals onto an atomically flat Au(111) surface by thermal sublimation of the molecular powder in ultrahigh vacuum at a temperature of 563 K. Scanning tunneling microscopy (STM) images at 4 K show that DIAn diradicals assemble in two types of structures for sub-monolayer coverages: extended close-packed domains and molecular chains (\Figref{FigTopoDomain}b). The former consists of a parallelogram network with a unit cell (white box in \Figref{FigTopoDomain}c) containing two DIAn molecules, as indicated in the molecular model in \Figref{FigTopoDomain}c. Differential conductance ($dI/dV$) spectra on the molecules in these close-packed domains show two bias-symmetric steps at $\pm$11 meV (\Figref{FigTopoDomain}d). The spectral feature is highly localized at the central anthracene core and absent over the protruding groups. We tentatively attribute it to the inelastic excitation of a molecular vibrational mode \cite{Stipe1998, Bachellier2020}. \np{In fact, Raman spectra of DIAn crystals show a prominent band around 11 meV corresponding to the out-of-phane deformation of the mesityl and of the isopropylsilyl groups (Figs. S1, S2).}
Molecular chains, in contrast to the domains, are rather disordered structures, weakly packed, and oriented along the Au(111) herringbone reconstruction.

To explore single DIAn diradicals, we moved apart individual molecules from the chain structures using the tip of the STM. As in the domains, DIAn molecules appear in STM images as two inverted protrusions, each composed of the bulkier mesityl groups (Mes) and the lower (triisopropylsilyl)ethynyl moiety (CCSi(i-Pr)$_3$), as shown in Figs.~\ref{FigTopoDomain}e-f. It is important to note that, in some occasions, tip-manipulated DIAn molecules appeared with an asymmetric configuration, \np{with one half appearing with lower height (top right of \Figref{FigTopoDomain}e).  Although we cannot state the precise origin of this different aspect, we speculate that it might be due to a distortion (e.g. a rotation) of the mesityl groups.}

\begin{figure*}[!ht]
   \includegraphics[width=0.80\textwidth]{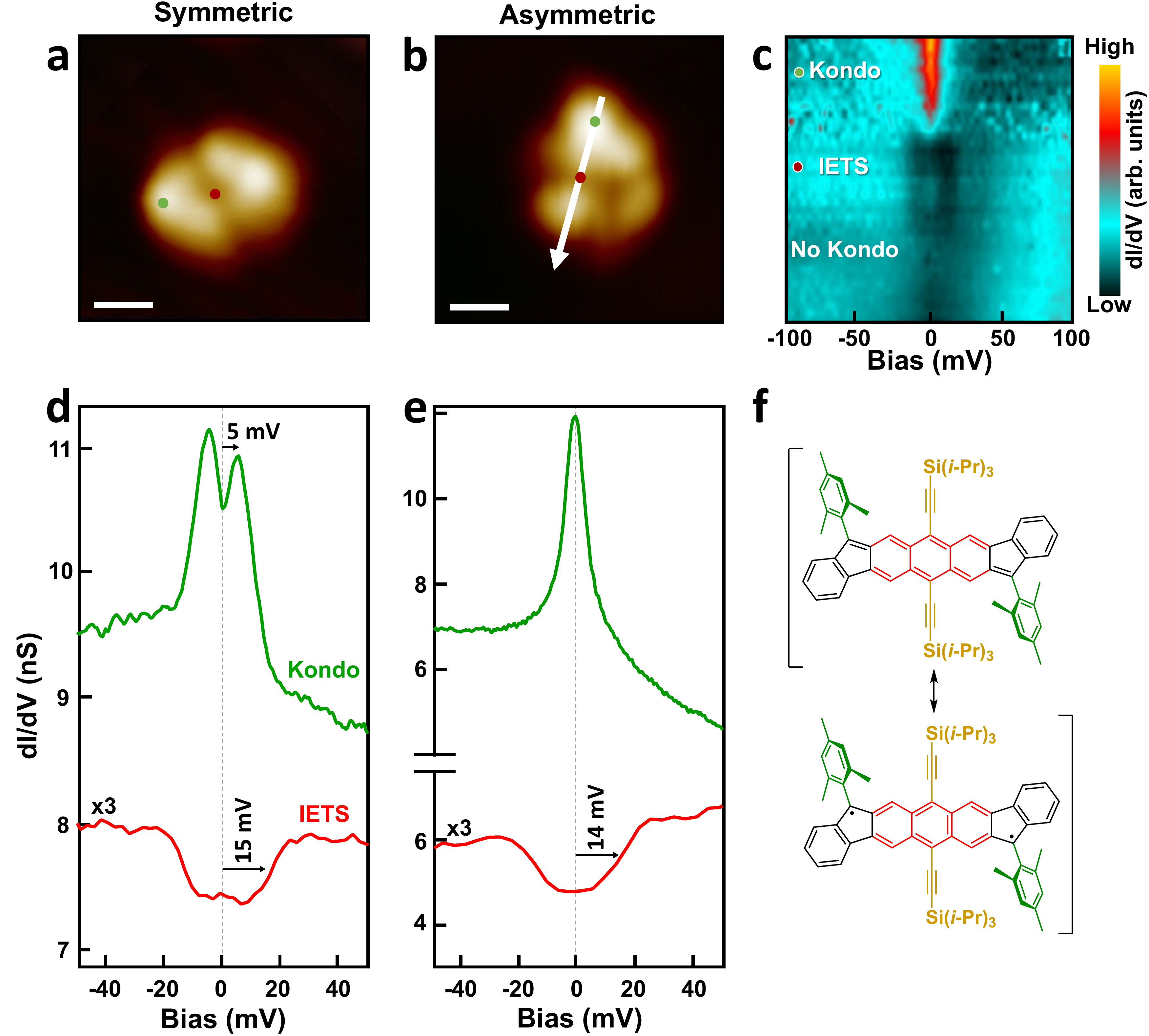}
	\caption{\label{FigDoubleKondo} 
	\textbf{Magnetic fingerprints of isolated DIAn diradical molecule.} (a-b) Topographic images of the symmetric and asymmetric DIAn diradical. The scale bar corresponds to 1 nm. Colored circles in the topographic image mark the positions where $dI/dV$ spectra were recorded. The corresponding $dI/dV$ spectra measured on symmetric and asymmetric DIAn diradical are plotted in (d,e), respectively. A split Kondo feature is observed at the mesityl groups (Mes), confirming the diradical character of the symmetric DIAn molecule, whereas a single Kondo peak is observed for the asymmetric DIAn molecule (green curves). In both cases, a vibrational inelastic electron tunneling (IETS) feature is observed on the central anthracene core of the molecule. The signal amplitude of the IETS spectra was multiplied by three for a better reading (red curves). (c) $dI/dV$ stacked plot taken along the white arrow in b), showing that the Kondo is located on the mesityl group of the molecule while the vibrational IETS feature is found at the anthracene position (red). The Kondo appears at only one side of the asymmetric molecule, on its brighter mesityl group. (f) Non-Kekulé resonance diradical structure of DIAn, showing that the two radicals are located at the junction between the indene (black) and mesityl (green) groups of the molecule. }
\end{figure*}

\np{The open-shell character of the isolated DIAn species could be resolved via STS measurements. Differential conductance ($dI/dV$) spectra over their mesityl groups (\Figref{FigDoubleKondo}a) resolved now two narrow peaks centered symmetrically around zero bias (green curve in \Figref{FigDoubleKondo}d), which reveal the survival of their diradical state on the Au(111) substrate. Owing to a weak interaction of the radical state with the metal surface, its associated spin S=1/2  can be partially screened via the Kondo effect \cite{Kondo1964, Ternes2009, Scott2010, Zhang2015, Hellerstedt2018}, resulting in a narrow logarithmic resonance at zero bias. However, over DIAn's bulky sides, we observed, instead, a double-peak structure. This is attributed to the presence of two antiferromagnetically interacting radical states.
As shown in \Figref{FigDoubleKondo}f, DIAn can lie either in an open shell non-Kekulé structure or in a closed shell state. In the former, two radicals appear localized at the junctions between the indene and mesityl groups (i.e., at the apical carbons of the five-membered rings), spaced by the central anthracene bridge, which mediates their antiferromagnetic interaction  $J$ between them, forming a singlet ground state. 
In this configuration, the lower dI/dV signal at zero-bias reflects the absence of Kondo states for the singlet ground state. At tunneling energies equal to or larger than $J$, dI/dV peaks reflect the excitation of a triplet state, with their corresponding energy providing a direct measure of $J$  \cite{Li2019,Mishra2020}. Above these onsets, Kondo-like dynamical fluctuations in the triplet state, now accessible by inelastic tunneling electrons, results in the logarithmic tail and the appearance of a split-Kondo resonance rather than inelastic steps \cite{Paaske2006, Ternes2015}. }
  
\np{In contrast, on the distorted DIAn molecules, a single dI/dV peak centered at zero bias is measured only over the brighter mesityl group (green curve, \Figref{FigDoubleKondo}e), while a featureless spectrum appears on the lower half of the molecule ($dI/dV$ stacked plot, \Figref{FigDoubleKondo}c). This reveals that on the \textit{distorted} side, the radical state is either quenched probably due to a conformational change or a strong interaction with the metal underneath. Additionally, bias-symmetric dI/dV steps at $\pm 15$~meV were detected over the central anthracene core on both pristine and distorted DIAn species (red curves in \Figref{FigDoubleKondo}d,e). Since we observe them with similar intensity on the diradical and monoradical species, we can safely exclude their magnetic origin. Instead,  we tentatively assign them to the excitation of the vibrational mode observed on the assembled domains, here at slightly higher energy probably due to the different conformation adopted when isolated on the metal substrate. }

The shifted Kondo resonances on DIAn molecules amount to a singlet-triplet excitation energy of $E_{ST}\simeq$ 5 meV. This value is significantly smaller than the singlet-triplet energy difference obtained from density functional theory (DFT) simulations of DIAn in the gas phase \cite{Rudebusch2016}. In particular, our DFT results in Figs. S3a and S3b corroborate the singlet ground state, and find a parallel spin configuration at almost 140 meV above, signaling for a singlet-triplet gap $E_{ST}$ much larger than in our experiment, and similar to previous measurements of DIAn \cite{Dressler2018}.

The origin of the smaller inter-radical exchange interaction observed in the experiment is likely connected with molecular modifications when adsorbed onto the gold surface. We have simulated the geometric relaxation and magnetic state of a DIAn molecule on a Au(111) slab using DFT (results summarized in Figs. S3-S7). The first important result is that DFT reproduces the survival of the diradical character of DIAn molecules on the gold surface (Fig. S3g). The molecule is physisorbed and keeps C-Au distances larger than 3 \AA. After geometry optimizations, two magnetic configurations were converged corresponding to a $S=0$ ground state and an $S=1$ ($2\mu_{B}$) excited state, with an energy spacing significantly smaller than in the gas phase. Thus, DFT confirms that, on the gold surface, the exchange interaction between indene radicals is reduced. The metallic substrate imposes, first, a partial planarization on the molecular structure by rotating the mesityl groups (Mes) towards the diindeno anthracene plane. As shown in Fig. S4, the mesityl rotation induces a slight delocalization of the radical towards their aromatic center,  which effectively increases the separation between spin centers and, consequently, reduces their magnetic interaction. Additional corrections to the simulations due to Coulomb correlations (considered in Fig. S6), or the renormalization of exchange coupling due to Kondo correlations \cite{jacob2021renormalization} can account for the further reduction of the reduced exchange observed in our experiments.

Our findings demonstrate that the commonly assumed inert role of molecular substituents \cite{Rudebusch2016} (e.g. Mes, CCSi(i-Pr)$_3$) on the active magnetic properties of diindeno[\textit{b,i}]anthracene derivatives needs to be revised when the diradicals are put in contact with a metallic electrode. Although, unpaired electron spins remain present in the molecule, their interaction is altered significantly upon adsorption on a metallic electrode. We found that the CCSi(i-Pr)$_3$ group connected at the anthracene core allows preserving the spin polarization of isolated DIAn diradicals on gold. Fig. S7 shows that when the two CCSi(i-Pr)$_3$ groups were replaced by CCSiMe$_3$, DIAn molecules are found to adsorb closer to the surface. In that case, the gold substrate imposes a bending of CCSi(i-Pr)$_3$ triple bonds (Fig. S7a-c), while a gold atom is slightly pushed up from the surface plane (Fig. S7a). Such modification of the molecule-substrate separation modifies the charge distribution and quenches spin polarization over the molecular backbone.   

\begin{figure*}[!ht]
   \includegraphics[width=0.75\textwidth]{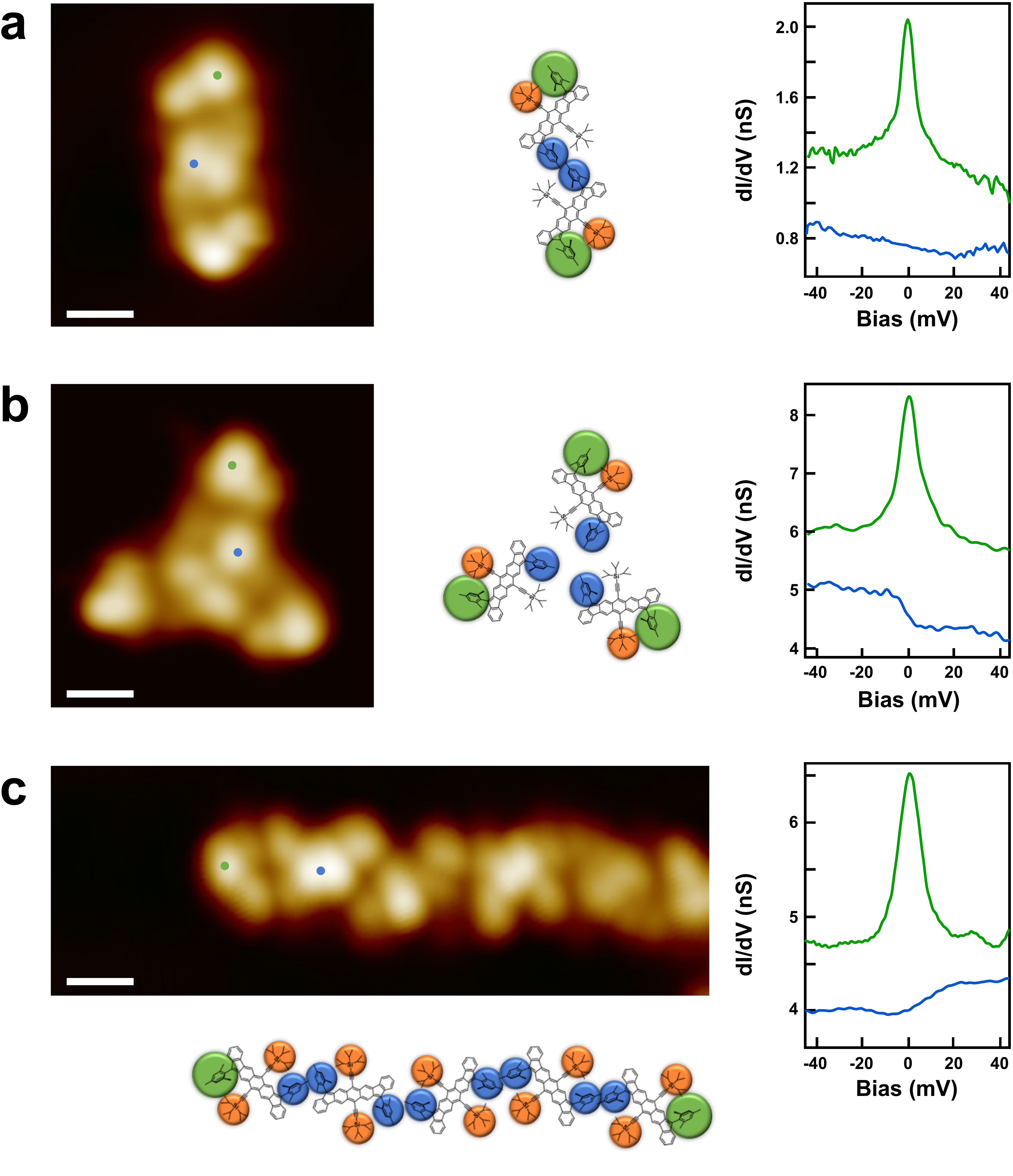}
	\caption{\label{FigKondoDimerTrimer} 
	\textbf{Kondo feature in nanostructures made of a few DIAn molecules.} STM images showing close-packed nanostructures made of several DIAn molecules, with the corresponding chemical structures and characteristic $dI/dV$ spectra taken over the regions highlighted in green and blue in the chemical models. The data were obtained for: (a) dimer; (b) trimer; and (c) molecular chain of DIAn molecules. Evident Kondo features are observed on specific positions of the nanostructures. Color code: strongly interacting mesityl group (blue), non-interacting mesityl group (green), triisopropylsilyl-ethynyl group (orange). Scale bars in a-c) correspond to 1 nm.}
\end{figure*}

Next, we show that the spin interaction in DIAn diradicals is modified in close-packed nanostructures made of a few molecules. Figure \ref{FigKondoDimerTrimer} shows nanostructures formed by two, three, or more DIAn molecules. Spectra recorded on the interacting mesityl groups of a DIAn dimer (Figure \ref{FigKondoDimerTrimer}a, blue dot), reveal the absence of a magnetic fingerprint (blue curves), while spectra on the non-interacting mesityl groups (green dot), reveal now a single Kondo feature (green curves). \np{This is further confirmed by spatial maps of the Kondo resonance shown in Fig. S8, which illustrate the localization of the  Kondo signal exclusively over external mesityl termini.}
This observation is consistent across other interacting molecular nanostructures, such as trimers (Figure \ref{FigKondoDimerTrimer}b) and molecular chains (Figure \ref{FigKondoDimerTrimer}c).  

\np{The absence of a Kondo resonance on the neighbouring mesityls can be explained caused as a combination of both a finite overlap of their electronic wavefunction of two neighboring spins and the concomitant intramolecular rotation of the involved mesityl groups for the assembly. Nevertheless, the quenching of one of the two radicals of DIAn also explains the observation of a single Kondo resonance on the external mesityl groups in DIAn dimers, trimers, and chains, in clear contrast with the split Kondo feature observed in isolated molecules.} This observation also explains the complete quenching of all Kondo signals in extended 2D close-packed domains, as shown earlier (\Figref{FigTopoDomain}).

In conclusion, here we report the survival of the magnetic state of DIAn diradicals upon their adsorption on a Au(111) surface. We have studied the spin interactions in the molecule through the many-body Kondo effect by performing scanning tunneling spectroscopy measurements and density functional theory calculations. The significant results from this study are (1) the persistence of the diradical singlet ground state upon adsorption; (2) the detection of spin interactions between the two spin centers of a single diradical; (3) the control over inter-molecular spin interactions through the formation of nanostructures by self-assembly or STM tip manipulation; and (4) the demonstration of the substrate influence on the lowering of the singlet-triplet gap of the DIAn diradical by imposing a rotation of its mesityl groups. Importantly, these are crucial questions raised that need to be solved for the future use of diradicals in spintronic and NLO devices, in which the molecules are contacted with metal electrodes. Our study paves the way to this by precise control of the spin interactions in isolated molecules and assembled nanostructures made of diradicals.

\section{Supporting Information}
The supporting information file is composed of the following:
\np{
\begin{enumerate}[label=\Alph*]
    \item Vibrational mode of DIAn,
    \item Energy Calculation of Singlet-Triplet States in DIAn adsorbed on gold,
    \item Change in Electronic Structure After Rotation of Mesityl Groups,
    \item Projected Density of States and Bond Order Modification in the Gas Phase,
    \item Influence of Coulomb Repulsion U in the HOMO-LUMO gap, Singlet-Triplet Energy and Bond Lengths,
    \item Influence of Substituent Chemical Composition on the Magnetic Properties of the DIAn Molecule,
    \item Kondo Map of DIAn Trimers,
    \item Controlling the Kondo Signature by Lateral Manipulation,
    \item Comparison of the long-range spectra of molecular chains and domains,
    \item Correlation between Molecular Orbital positions and Kondo Resonance,
    \item Methods.
\end{enumerate}
}

\subsection{Notes}
The authors declare no competing financial interest.

%\section*{Acknowledgments}
\begin{acknowledgement}
We gratefully acknowledge financial support from Spanish MCIN/AEI/ 10.13039/501100011033 and the European Regional Development Fund (ERDF) through grants with number PID2019-107338RBC61, PID2020-115406GB-I00, PID2021-127127NB-I00, and CEX2020-001038-M, and from the European Union (EU) through Horizon 2020 (FET-Open project SPRING Grant.~no.~863098).  J.C.~also acknowledges the Junta de Andalucía, Spain (PROYEXCEL-0328), and M.M.H.~acknowledges the NSF (CHE-1954389) for financial support. 
\end{acknowledgement}

\bibliography{references.bib}

\end{document}